\documentclass[
    ,final            
  ]
  {aipproc}

\layoutstyle{6x9}

\begin{document}

\title{QCD corrections to the electroproduction of hadrons with high $p_T$}


\author{A. Daleo}{ address={Institut f\"ur Theoretische Physik,  
Universit\"at Z\"urich, Winterthurerstrasse 190, CH-8057 Z\"urich, Switzerland
}}

\author{D. de Florian}{ address={Departamento de F\'{\i}sica,
Universidad de Buenos Aires, Ciudad Universitaria, Pab.1 (1428)
Buenos Aires, Argentina}}

\author{R. Sassot}{
  address={Departamento de F\'{\i}sica,
Universidad de Buenos Aires, Ciudad Universitaria, Pab.1 (1428)
Buenos Aires, Argentina}}

\begin{abstract}
We compute the order $\alpha_s^2$ corrections to the one particle inclusive
electroproduction cross section of hadrons with non vanishing transverse 
momentum. We compare our results with  H1 data on forward production of 
$\pi^0$, and conclude that the data is well described by the DGLAP approach. 
 within the theoretical uncertainties. 
\end{abstract}

\maketitle

\section{Introduction}

The precise measurement of final state hadrons in lepton nucleon deep 
inelastic scattering constitutes an excellent benchmark for different 
features of perturbative quantum chromodynamics. Among them,
the calculation of higher order corrections, which have been 
explored and validated for most processes up to next to leading order (NLO) 
accuracy. For the one particle inclusive processes only very recently there 
has been progress beyond the leading order (LO) 
\cite{gluon,quark,aure,Fontannaz:2004ev,Maniatis:2004xk}. 
and up to now there were no analytic computation of the 
${\cal{O}}(\alpha_s^2)$ corrections for the 
electroproduction of hadrons with non vanishing transverse momentum. The 
analytic computation of the ${\cal{O}}(\alpha_s^2)$ corrections allows us to 
check factorization in a direct way, which means that collinear 
singularities showing up in the partonic cross section
factorize into parton densities (PDFs) and fragmentation functions (FFs).
As a consequence of this explicit cancellation, the resulting cross section
is finite and can be straightforwardly convoluted with PDFs and FFs in a fast
and stable numerical codes. The analytical result is still sufficiently 
exclusive and keeps the dependence 
on the rapidity and the transverse momentum of the produced hadron, allowing 
a detailed comparison with the experimental data.
In the following we summarize the results obtained in Ref. \cite{Daleo:2004pn}

\section{${\cal{O}}(\alpha_s^2)$ QCD corrections}
 We consider the process 
\begin{equation}
l(l)+P(P)\longrightarrow l^{\prime}(l^\prime)+h(P_h)+X,
\end{equation}
where a lepton of momentum $l$ scatters off a nucleon of momentum $P$ with 
a lepton of momentum $l^{\prime}$ and a hadron $h$ of momentum $P_h$ tagged 
in the final state. Omitting target fragmentation at zero transverse momentum, 
which has been discussed at length in \cite{gluon,quark}, the cross 
section for this process can be written as 
\begin{equation}\label{eq:dsigma}
\frac{d\sigma^{h}}{dx_B\, dQ^2}=\sum_{i,j,n}\,\int_{0}^{1}d\xi\,\int_{0}^{1}d\zeta\,\int 
\mbox{dPS}^{(n)}\,\left[f_i(\xi)\,D_{h/j}(\zeta)\,\frac{
d\sigma^{(n)}_{ij}}{dx_{B}\,dQ^2\,\mbox{dPS}^{(n)}}\right]
\end{equation}
where $\sigma^{(n)}_{ij}$ is the partonic level cross section 
corresponding to the process and is calculated order by order 
in perturbation theory through the related parton-photon squared matrix 
elements 
$\overline{H}^{(n)}_{\mu\nu}(i,j)$ for the $i+\gamma\rightarrow j+X$ processes
\begin{equation}\label{eq:psigma}
\frac{d\sigma^{(n)}_{ij}}{dx_B\, dQ^2\,dy\,dz}=\frac{\alpha_{em}^2}{e^2}
        \frac{1}{\xi\,x_B^2\,S_H^2}
        \left(Y_M (-g^{\mu\nu})+Y_L \frac{4x_B^2}{Q^2}P^{\mu}P^{\nu}
        \right)\sum_{n}\overline{H}^{(n)}_{\mu\nu}(i,j)\, .
\end{equation} 
in terms of the standard kinematical variables \cite{Daleo:2004pn}.

At order-$\alpha_s^2$, the partonic cross sections receive contributions from 
the following  reactions:
\begin{equation}\label{eq:partreac}
\begin{array}{ll}
\mbox{Real contributions}&\left\{
\begin{array}{lcl}
\gamma+q(\bar{q})&\rightarrow& g+g+q(\bar{q})\\
\gamma+q_{i}(\bar{q_{i}})&\rightarrow&q_{i}(\bar{q_{i}})+q_{j}+\bar{q_{j}}
\,\,\,\,(i\neq j)\\
 \gamma+q_{i}(\bar{q_{i}})&\rightarrow&q_{i}(\bar{q_{i}})+q_{i}+\bar{q_{i}}\\
\gamma+g&\rightarrow& g+q+\bar{q}
\end{array}
\right.\\
\mbox{Virtual contributions}&\left\{
\begin{array}{lcl}
\gamma+q(\bar{q})&\rightarrow& g+ q(\bar{q})\\
\gamma+g&\rightarrow& q+\bar{q}
\end{array}
\right.
\end{array}
\end{equation}
where any of the outgoing partons can fragment into the final state hadron 
$h$.

At variance with the $|p_T|=0$ case, where the integration 
over final states leads to overlapping singularities along various curves in 
the residual phase space, here the only remaining singularities are found at 
$z=0$ and thus they can be dealt with the standard method. 
After combining real and virtual contributions to a given partonic process, 
the cross section can be written as 
\begin{eqnarray}\label{eq:psigma2}
\frac{d\sigma^{(2)}_{ij}}{dx_B\,dQ^2\,dy\,dz}&=& \frac{c_q\,C_{\epsilon}^2}
{\xi\,x_B^2\,S_H^2}\,
\left\{\frac{1}{\epsilon}\,{\cal{P}}^{(2)}_{1\, ij}(\varrho,y,z)\,
        +C^{(2)}_{ij}(\varrho,y,z)+{\cal O}(\epsilon)
        \right\}
\,, 
\end{eqnarray}
where the coefficient of the single poles, 
${\cal{P}}^{(2)}_{1\,ij}(\varrho,y,z)$, as well
as the finite contributions $C^{(2)}_{ij}(\varrho,y,z)$, include `delta' and 
`plus' distributions in $z$.
The IR double poles present in the individual real and virtual contributions 
cancel out in the sum, providing the first straightforward check on the angular
integration of real amplitudes and the loop integrals in the virtual case. 
In the real terms, the above 
mentioned double poles come from the product of a pole arising in the 
integration over the spectators phase space  and a single pole coming from the 
expansion of $z^{-1+\epsilon}$ factors. Double poles in the virtual 
contributions always arise from loop integrals.

The remaining singularities, contributing to the single pole, are of UV and 
collinear origin. The former are removed by means of coupling constant 
renormalization, whereas the latter have to be factorized in the redefinition 
of parton densities and fragmentation functions.

\section{Phenomenology}

\begin{figure}[h]
\includegraphics[width=7cm]{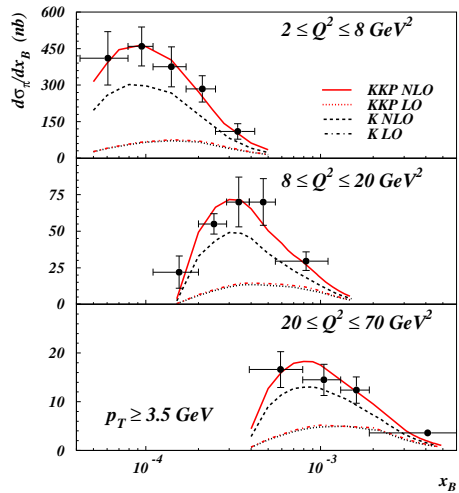}
\caption{LO and NLO cross sections, including experimental cuts as explained in the text,
as a function of $x_B$. H1 data \cite{Aktas:2004rb} for the range $p_T\ge 3.5\mbox{ GeV}$
are also shown.
}\label{fig:x_Bj}
\end{figure}

In Figure \ref{fig:x_Bj} we show the LO and NLO 
predictions for the electroproduction of neutral pions as a function of 
$x_B$ and $p_T$, respectively, in the kinematical range of the H1 
experiment  \cite{Aktas:2004rb}, together with the most recent data for 
the range $p_T\ge 3.5\mbox{ GeV}$. 
The cross sections are computed as described in the previous sections, 
applying H1 cuts and using MRST02 parton densities \cite{MRST02}.
Similar results are found using other sets of modern PDFs. For the 
input fragmentation functions, we use two different sets, the ones from 
reference \cite{KKP} denoted as KKP and those from \cite{kretzer} referenced 
as K. We set the renormalization and factorization scales as the average between $Q^2$ and $p_T^2$, and 
we compute $\alpha_s$ at NLO(LO)  fixing $\Lambda_{QCD}$ as in the MRST 
analysis.
\setlength{\unitlength}{1.mm}
The plots clearly show that the NLO cross sections are much larger than the 
LO ones, even by the required order of magnitude in certain kinematical 
regions. Another interesting feature is that the uncertainty due to the choice 
of a fragmentation functions set is also quite noticeable, this fact driven by 
the different gluon content of the two sets considered here. Low $Q^2$ bins 
seem to prefer KKP set, which have a larger gluon-fragmentation content, 
whereas for larger $Q^2$ both sets agree with the data within errors. LO 
estimates show a much smaller sensitivity on the choice of fragmentation 
functions, since gluon fragmentation does not contribute significantly to 
the cross section at this order.

The rather large size of the K-factor can, then, be understood  
as a consequence of the opening of a new dominant (`leading-order') channel,
 and not to the `genuine' increase in the partonic cross 
section that might otherwise threaten perturbative stability. The dominance of
the new channel is due to the size of the gluon distribution 
at small $x_B$ and to the fact that the H1 selection cuts highlight
the kinematical  region dominated by the $\gamma+g\rightarrow g+q+\bar{q}$ 
partonic process.

\setlength{\unitlength}{1.mm}
\begin{figure}[hbt]
\includegraphics[width=7cm]{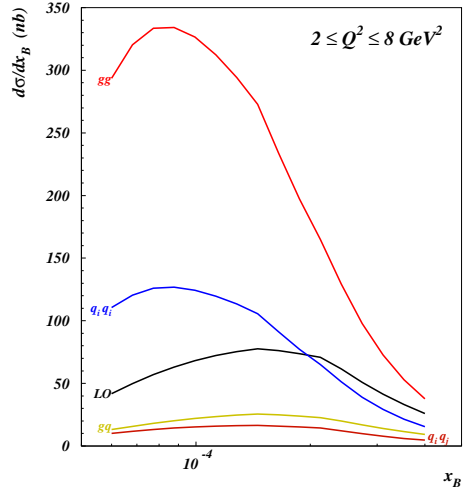}
\caption{Contributions to the cross section  for the lowest $Q^2$ bin of Figure \ref{fig:x_Bj}}
\label{fig:discri}
\end{figure}
In Figure \ref{fig:discri}  we show the different contributions to the cross 
section discriminated by the underlying partonic process. Notice that at very 
small $x_B$ the $gg$ term can be by itself several times larger than the LO 
contribution, remaining larger or comparable even for higher $x_B$ values. 
The forward selection is also responsible of the scale sensitivity of the
cross section, as it supresses large components  with small scale
dependence whereas it stresses components as $gg$ whose scale dependence would
be partly canceled only at NNLO

\begin{theacknowledgments}
Partially supported by CONICET, Antorchas,  UBACYT and ANPCyT, Argentina,
and the Swiss National Science Foundation (SNF) through grant No. 200021-101874.

\end{theacknowledgments}

\end{document}